# Hybrid Single-Pulse and Sawyer-Tower Method for Accurate Transistor Loss Separation in High-Frequency High-Efficiency Power Converters

Xiaoyang Tian, Mowei Lu, Florin Udrea, and Stefan M. Goetz

*Abstract*—Accurate measurement of transistor parasitic capacitance and its associated energy losses is critical for evaluating device performance, particularly in high-frequency and high-efficiency power conversion systems. This paper proposes a hybrid single-pulse and Sawyer–Tower test method to analyse switching characteristics of field-effect transistors (FET), which not only eliminates overlap losses but also mitigates the effects of current backflow observed in traditional double-pulse testing. Through a precise loss separation model, it enables an accurate quantification of switching losses and provides a refined understanding of device energy dissipation mechanisms. We validate the hysteresis data and loss separation results through experimental measurements on a 350-W *LLC* converter, which further offers deeper insights into FET dynamic behaviour and its dependence on operating conditions. This method is applicable to a wide range of FETs, including emerging SiC and GaN devices, and serves as a valuable tool for device characterization and optimization in power electronics.

## I. INTRODUCTION

THE rapid advancement of renewable energy systems, electric vehicles, and industrial automation has massively stimulated the demand for high-frequency, high-efficiency power converters in modern power electronics [1–7]. The concurrent growth of available devices requires better understanding of the switching characteristics of wide-bandgap semiconductors, including silicon (Si), silicon carbide (SiC), and gallium nitride (GaN), for device selection, system performance optimisation, and reliability analysis [8, 9, 10]. However, these devices exhibit unique dynamic characteristics, such as nonlinear output capacitance and hysteretic charge–discharge behaviour [11], which significantly impact switching performance and energy losses. Optimisation of device operation and circuit-level design requires accurate characterisation of these properties and isolation of the associated energy loss components.

Existing methods for the analysis of switching characteristics often involve complex experimental setups and expensive equipment, such as source measurement units (SMUs) or capacitance-measurement units (CMUs), which present significant challenges for both research and industrial applications. Moreover, while most of these methods focus on measuring total switching losses, they fail to provide a detailed breakdown of the distinct contributions to these losses, such as the separation of transistor overlap energy losses from turn-off energy losses, even under soft-switching conditions [12, 13]. For example, the Sawyer–Tower (ST) method [14] requires complex setups involving external capacitors and high-voltage probes. Yet it lacks the flexibility to precisely control key operating conditions, such as voltage and frequency. On the other hand, the widely used double-pulse test (DPT) [12] is susceptible to inaccuracies due to load-current backflow and overlap losses, which complicate isolating charge and discharge energy components. These limitations not only compromise the efficiency and cost-effectiveness of device testing but also reduce the accuracy of performance evaluations, particularly for high-performance devices.

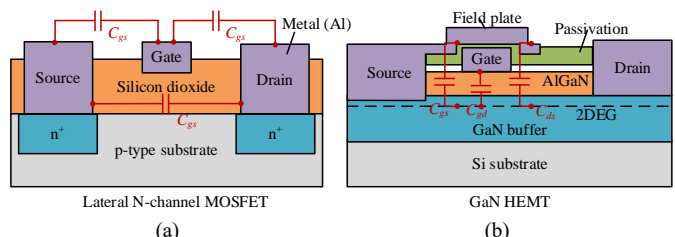

Fig 1. Simplified structure of lateral (a) N-channel MOSFET and (b) GaN HEMT.

To solve these challenges, this paper presents a single-pulse platform based on a combined setup including a half-bridge inverter and a signal-generator–power-amplifier test base, designed to provide efficient, accurate, and cost-effective switching loss analysis and separation. We practically tested and validated the loss analysis and hysteresis data have on a 400-V, 250-kHz, 350-W *LLC* DC-DC converter.

Section II introduces the proposed combined testing platform and operation principle. Section III explores an accurate loss separation model, which is further validated in a high-frequency *LLC* converter equipped with three distinct alternative Si, SiC, and GaN inverters. Section IV further demonstrates the accuracy and practicality of the proposed method through measurement data. Finally, Section V draws conclusions.

## II. PROPOSED BENCHMARKING PLATFORM

Figure 1 illustrates a simplified lateral N-channel power MOSFET and a GaN HEMT. The output capacitance, denoted as $C_{oss}$, comprises the drain–source capacitance ($C_{ds}$) and gate–source capacitance ($C_{gs}$). This capacitance determines a large share of the switching losses. It furthermore affects the resonant frequency and its hysteresis properties, which entail additional switching losses.

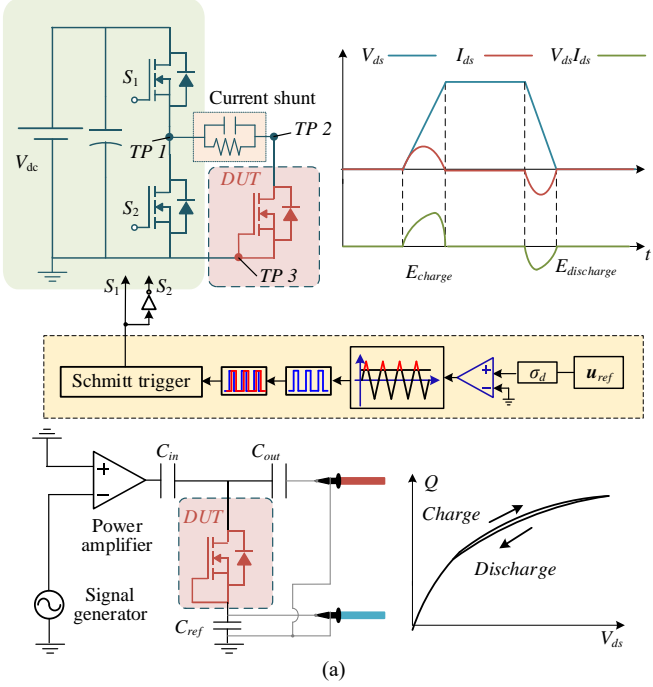

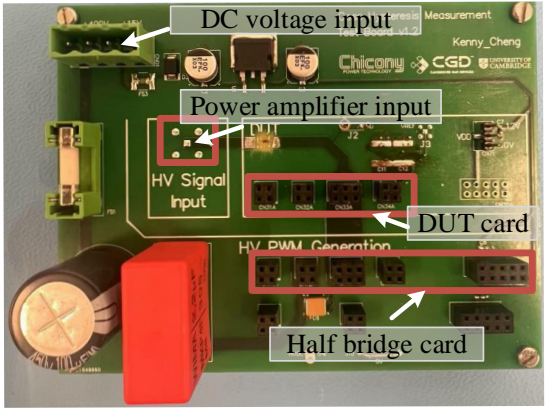

(b)
Fig 3. Proposed benchmarking platform. (a) Circuit topology. (b) Integrated motherboard design.

We combine a half-bridge-based single-pulse test and the ST setup (Fig. 2) for an accurate analysis. The gate–source terminals of the device under test (DUT) are shorted to ensure that it remains in the off-state. The gate–source short allows to model the DUT as an anti-parallel diode and an output capacitor. Under these conditions, a single pulse serves to accurately record the instantaneous voltage and current values (which can be measured through the current shunt) during the pulse. During the turn-off transient, the drain current charges the output capacitance of the MOSFET. As a result, the measured energy loss corresponds to the energy stored in the output capacitor. This method remains simple and effective across a wide range of input voltages and frequencies. Additionally, this method eliminates the reverse recovery losses of the diode. This simulates the switching characteristics of the device under soft-switching conditions. As a result, the turn-on and turn-off losses are measured with the half-bridge circuit, whereas the charging and discharging behaviour is characterized with the ST setup. Furthermore, voltage and current waveforms of the half-bridge setup can also inform the charging and discharging characteristics, where the current-over-time integration delivers the charge values. Therefore, comparative evaluation of these datasets can accurately determine the energy overlap, charging as well as discharging energy, and the $C_{oss}$ hysteresis curve.

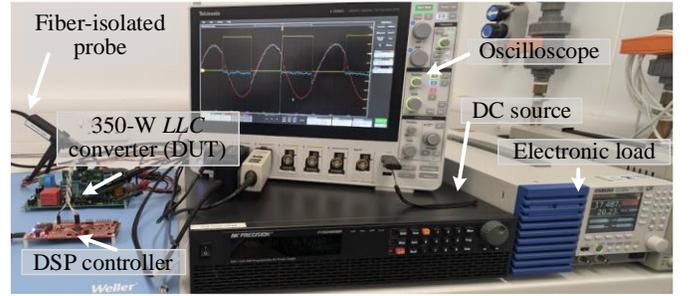

Fig. 4. Test base for GaN-based *LLC* converter.

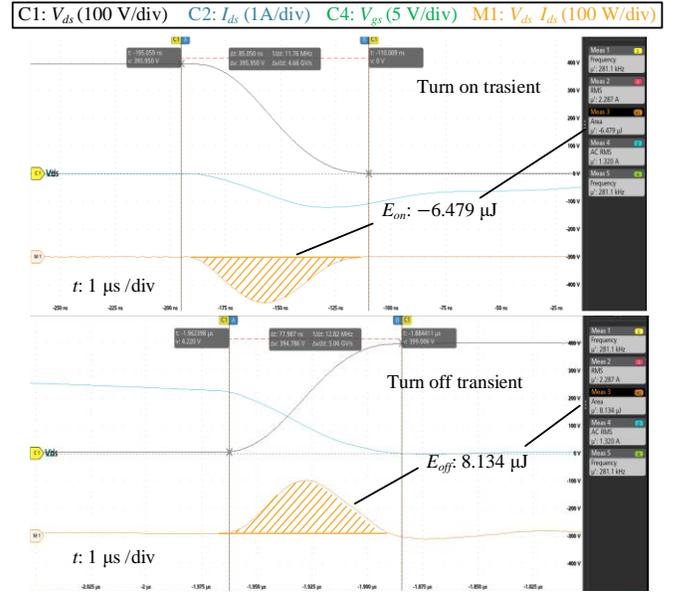

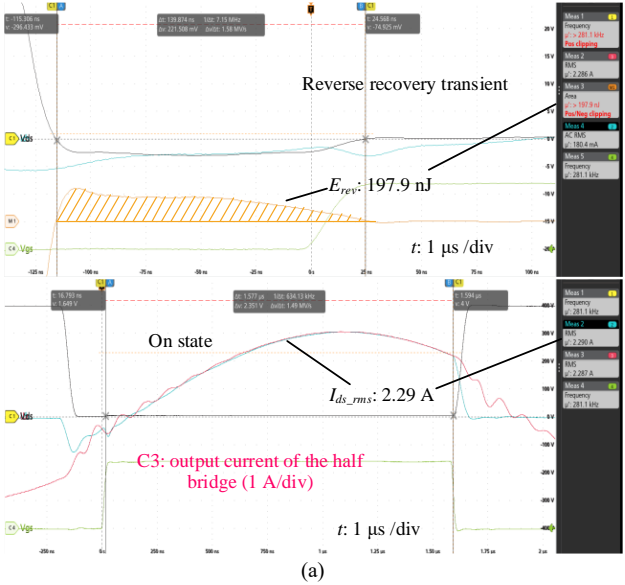

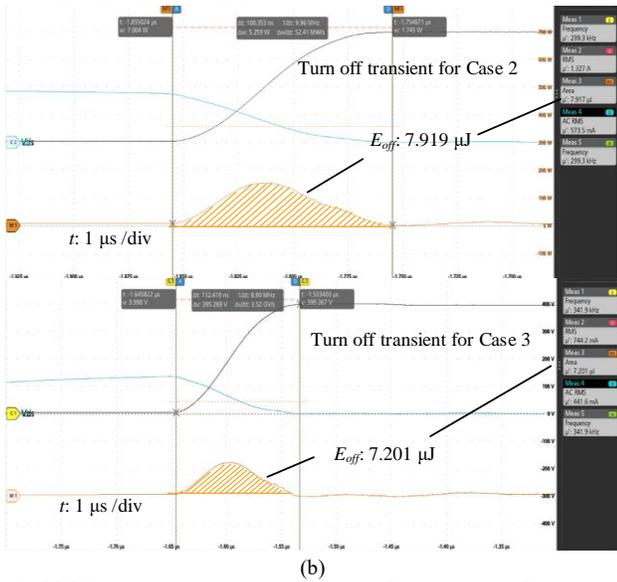

Fig 5. *LLC* loss component measurement. (a) Turn-on, turn-off, reverse and conduction energy losses: Case 1: 17.5 A, 281.1 kHz. (b) Turn-off transient energy losses under different load conditions: Case 2: 8.75 A, 299.3 kHz; Case 3: 1.75 A, 341.9 kHz.

TABLE I
PARAMETERS OF GaN-BASED *LLC* CONVERTER

| Items | Value |
|---|---|
| GaN device | CGD65A055SH2 |
| Rated voltage (Input/Output) | 400 V/20V |
| Rated output current | 17.5 A |
| Rated output power | 350 W |
| Operation frequency | 250 – 350 kHz |

### III. LOSSES SEPARATION MODEL AND VALIDATION IN HIGH-FREQUENCY CONVERTERS

The switching losses of power MOSFETs can be calculated through

$$\begin{cases} E_{on} = E_{discharge} \\ E_{off} = E_{charge} + E_{overlap} \end{cases}, \quad (1)$$

where $E_{on}$ and $E_{off}$ denote the absolute value of the energy losses during turn-on and turn-off switching. The turn-on energy only includes the discharging energy of $C_{oss}$, while the turn-off energy includes its charging energy and the energy loss caused by the simultaneous nonzero product of the rising voltage and decaying current. This part of loss can be denoted by $E_{overlap}$ and calculated through

$$E_{overlap} = \int_{t_{off}} V_{ds}(t) \cdot I_{ds\_f}(t) \, dt, \quad (2)$$

where $V_{ds}$ represents the drain–source voltage of the MOSFET and $I_{ds\_f}$ the drain current. We can measure its value directly through the current shunt. The duration of the transient turn-off event is called $t_{off}$.

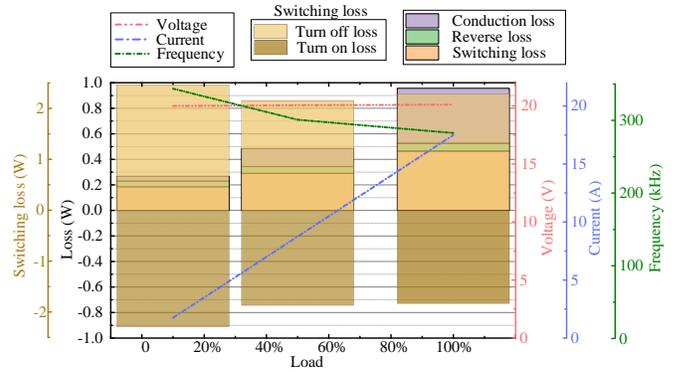

Fig 6. *LLC* loss separation for different cases.

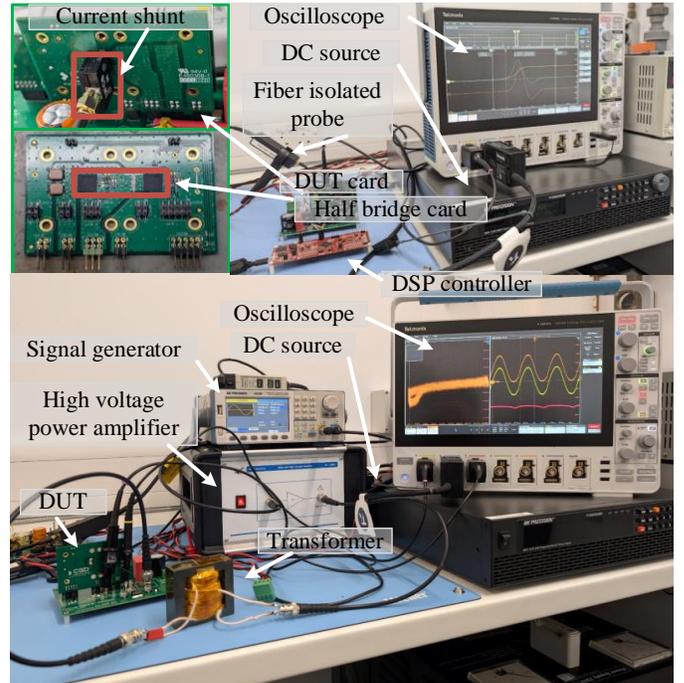

Fig. 7. Proposed hybrid single-pulse and ST benchmarking platform.

As discussed above, the main task in separating the loss components lies in the transient turn-off analysis. In practice, it is usually difficult to quantify the current share ($I_{ds}$) that charges the output capacitance and the one that is dissipated as heat. However, in some GaN transistor devices, the absence (or negligibility) of hysteresis offers an opportunity to isolate and analyse specific loss mechanisms. Specifically, under soft-switching conditions, the difference between the charging and discharging energy losses will be dominated by the overlapping energy losses per

$$E_{off} - E_{on} = E_{charge} - E_{discharge} + E_{overlap}$$
$$= E_{hysteresis} + E_{overlap} \quad (3)$$

We implemented a test setup for a GaN *LLC* converter (Fig. 4) and measured results for three load conditions under soft-switching operation (Fig. 5). Table I lists the basic parameters of the DUT. Figure 6 illustrates its loss characteristics for the three different operation cases.

## IV. NUMERICAL VERIFICATION OF THE PROPOSED METHOD

Figure 7 illustrates the practical benchmarking of the proposed hybrid method. We paired a high-voltage power amplifier with a transformer, which can turn the input signal into simultaneous high frequency and high voltage for the capacitance measurement. Table II lists the parameters of the testing devices and measuring equipment. Figure 8 (a) graphs the turn-on and turn-off waveforms of the GaN device under 400 V, whereas Fig. 8 (b) and (c) respectively present the voltage-dependent loss measurement of the device and the $C_{oss}$ hysteresis.

The turn-off energy loss of the DUT tested from the proposed method reaches 6.889 μJ, whereas the $C_{oss}$ hysteresis loss is 0.189 μJ. Therefore, based on the loss separation model, the loss components can be summarized and calculated as shown in Table III. Compared with the measured results of the *LLC* under 10% load condition, with the same GaN device, the error of the measured loss components for the DUT is only 0.034 μJ. Considering the differences in operating conditions, the error is within an acceptable range.

TABLE II
TEST BASE PARAMETERS

| Items | Value |
|---|---|
| Digital signal processer | TMS320F28379 |
| DUT (GaN) | CGD65A055SH2 |
| Typical $C_{oss}$ of DUT | 47 pF (at 400 V) |
| Current shunt | 10.4 mV/A |
| Reference capacitor | NP0 MLCC (500 V / 10 nF) |
| DC source | BK PVS60085MR (600 V/3 kW) |
| Function generator | BK 4062B (40 MHz) |
| Power amplifier | WMA-300 (±150 V / DC–5 MHz) |
| Passive voltage probe | 800 MHz / 1kV |
| Current probe | 120 MHz / 30 A |
| Fiber isolated probe | TIVP1 (1 GHz) |
| Oscilloscope | Tektronix MSO46 (12 bits / 1 GHz / 6.25 Gs) |
| Pulse duration | 2 μs (250 kHz) |

TABLE III
LOSS SEPARATION RESULTS

| Case | $E_{on}$ (μJ) | $E_{off}$ (μJ) | $E_{charge}$ (μJ) | $E_{overlap}$ (μJ) | $E_{hysteresis}$ (μJ) |
|---|---|---|---|---|---|
| | LLC measurement | | Half-bridge setup | Calculated through (1) | Calculated through (3) |
| 1 | 6.479 | 8.134 | 6.889 | 1.245 | 0.410 |
| 2 | 6.212 | 7.917 | | 1.028 | 0.677 |
| 3 | 6.658 | 7.201 | | 0.312 | 0.231 |

## V. CONCLUSION

This study presents a systematic and cost-effective approach for accurately separating FET switching losses by integrating a half-bridge-based single pulse test with the ST method. The single pulse test quantifies turn-on and turn-off losses of the DUT under different operation conditions. In combination with the loss separation model, the single pulse test quantifies the turn-on and turn-off losses of the DUT under various operating conditions. With the loss separation model, we can further compare the measured results with the $C_{oss}$ hysteresis curve and evaluate the discrete loss components. The measurement error in the experiments was within 40 nJ. This methodology allows for detailed quantification of overlap energy, charging/discharging energy, and dynamic behaviour and offer deeper insights into FET switching losses with low hardware complexity.

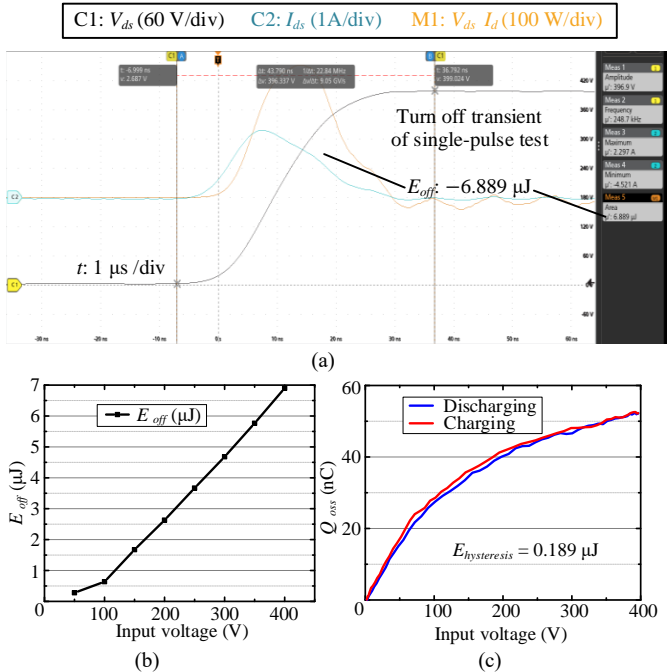

Fig. 8. Measured results of GaN device. (a) Waveforms under 400-V rated voltage. (b) Charging energy losses versus different $V_{ds}$. (c) $C_{oss}$ hysteresis.